\documentclass[12pt]{article}
\usepackage{epsfig}
\begin{document}

\title{PT and non-PT-Symmetric Solutions of the Schr\"odinger Equation for the Generalized Woods-Saxon Potential}
\author{C\"uneyt Berkdemir $^a$$\thanks{berkdemir@erciyes.edu.tr}$, Ay\c se Berkdemir $^a$$\thanks{arsland@erciyes.edu.tr}$ and Ramazan Sever
$^{b}$$\thanks{sever@metu.edu.tr}$\\
{\small {\sl $^a$ Department of Physics, Faculty of
Arts and Sciences, Erciyes University,}} {\small {\sl 38039, Kayseri, Turkey}}\\
{\small {\sl $^b$ Department of Physics, Middle East Technical
University,}} {\small {\sl 06531, Ankara, Turkey}}}

\date{\today}

\maketitle

\begin{abstract}

We investigate complex PT and non-PT-symmetric forms of the
generalized Woods-Saxon potential. We also look for exact
solutions of the Schr\"odinger equation for the PT and/or
non-PT-symmetric potentials of the kind mentioned above.
Nikiforov-Uvarov method is used to obtain their energy eigenvalues
and associated eigenfunctions.
\end{abstract}

\baselineskip=22pt plus 1pt minus 1pt

\vspace{0.5cm}


\maketitle

\section{Introduction}

A large variety of potentials with the real or complex forms are
encountered in various fields of the physics. A consistent
physical theory of quantum mechanics in terms of Hermitian
Hamiltonians is constructed on a complex Hamiltonian. In this
case, its energy levels are real and positive as a consequence of
$PT$-symmetry. Where $P$ and $T$ stand for the parity (or space)
and time reversal operators, respectively. It is also well-known
that $PT$-symmetry does not leads to completely real spectrums,
because there are several potentials where part or all of the
energy spectrums are complex. Exact solution of the Schr\"odiger
equation for these potentials are generally of interest \cite
{ref1, ref2, ref3}. Recently, Bender and his co-workers have
studied a number of complex potentials on $PT$-symmetric quantum
mechanics. They have showed that the energy eigenvalues of the
Schr\"odinger equation are real when $PT$-symmetry is unbroken,
whereas they come in the shape of complex conjugate pairs when
$PT$-symmetry is spontaneously broken \cite {ref4}. In these
studies, some numerical and analytical techniques have been used
to investigate non-Hermitian Hamiltonians with real or complex
spectra \cite {ref5, ref6}.

$PT$ invariant operators have been analyzed for real and complex
spectra by using a variety of techniques such as variational
methods \cite {ref8}, numerical approaches \cite {ref9},
semiclassical estimates \cite {ref10}, fourier analysis \cite
{ref11} and group theoretical approach with the Lie algebra \cite
{ref12}. It is pointed out that $PT$ invariant complex-valued
operators may have real or complex energy eigenvalues. Many
authors have studied on $PT$-symmetric and non-$PT$-symmetric
non-Hermitian potential cases such as flat and step potentials
with the framework of SUSYQM \cite {ref13}, exponential type
potentials \cite {ref14}, quasi exactly solvable potentials \cite
{ref15}, complex H\'{e}non-Heiles potential \cite {ref16} and deep
potential to describe optic-model analysis of elastic and
inelastic scattering processes \cite {ref17}.

Recently, an alternative method which is known as the
Nikiforov-Uvarov (NU) has been introduced in solving the
Schr\"{o}dinger equation. The solution of the Schr\"{o}dinger
equation for well-known potentials and Schr\"{o}dinger-like (i.e.
Dirac and Klein-Gordon) equations for a Coulomb potential have
been obtained by using this method \cite {ref18}. The increasing
interest on this solution method shows that $PT$-symmetric
potentials are also suitable in solving the Schr\"{o}dinger
equation for a exponential-type potential. This type potential has
been used in the work of Berkdemir {\it et.al.,} \cite {ref19} for
the nuclear scattering applications. The potential known as
Woods-Saxon has been generalized with an additional derivative
term and solved by using the NU method to obtain eigenvalue
equations. However, $PT$ and non $PT$-symmetric versions of the
generalized Woods-Saxon potential have not been introduced in the
literature and solved analytically for the Schr\"{o}dinger
equation. From this point of view, this method avoids direct
solving of Schr\"{o}dinger equation for these potential forms and
makes the problem more interesting.

The organization of the paper is as follows. In Sec. II, we
introduced a short summarization of the NU method, since the
detailed description has been already given in Ref.\cite {ref18}
and also used in Ref.\cite {ref19}. In Sec. III, the method
utilized to solve the Schr\"{o}dinger equation for $PT$ and
non-$PT$-symmetric non-Hermitian forms of the generalized
Woods-Saxon potential. Finally, in Sec. IV, we concluded the
article with a summary of our main remarks.

\section{Overview of the Nikiforov-Uvarov Method}
The method of solving second-order differential equations
developed by Nikiforov-Uvarov \cite {ref18} has been used in
appearing many problems of quantum mechanics. In principle, the
method makes it possible to present the theory of special
orthogonal functions \cite {ref21} by following a differential
equation of the form
\begin{equation}
\label{eq1} \psi ^{\prime \prime }(s)+\frac{\stackrel{\sim }{\tau
}(s)}{\sigma (s)}\psi ^{\prime }(s)+\frac{\stackrel{\sim }{\sigma
}(s)}{\sigma ^{2}(s)}\psi (s)=0,
\end{equation}
where $\sigma (s)$ and $\stackrel{\sim }{\sigma }(s)$ are
polynomials, at most second-degree, and $\stackrel{\sim }{\tau
}(s)$ is a first-degree polynomial. Following the method used in
Ref.\cite {ref19}, we transform the equation for $\psi (s)$ to an
equation of hypergeometric type
\begin{equation} \label{eq3}
\sigma(s)y^{\prime \prime }+\tau(s)y^{\prime }+\lambda y=0,
\end{equation}
by inserting $\psi(s)=\phi(s)y(s)$, where $\phi(s)$ satisfies the
equation $~\phi (s)^{\prime }/\phi (s)=\pi (s)/\sigma (s)$. $\tau
(s)=\stackrel{\sim }{\tau }(s)+2\pi (s)$ and its derivative has to
be negative. In the present case, the polynomial $\pi(s)$ is
\begin{equation}
\label{eq6} \pi =\frac{\sigma ^{\prime }(s)-\stackrel{\sim
}{\tau}(s)}{2}\pm
\sqrt{\left(\frac{\sigma^{\prime}(s)-\stackrel{\sim}{\tau}(s)}{2}\right)^{2}-
\stackrel{\sim}{\sigma}(s)+k{\sigma}(s)}
\end{equation}
where the parameter $k$ is a constant ($k=\lambda-\pi^{\prime }$).
$y(s)$ is the hypergeometric type function whose polynomial
solutions are given by Rodrigues relation
\begin{equation}
\label{eq4}
y_{n}(s)=\frac{B_{n}}{\rho (s)}\frac{d^{n}}{ds^{n}}\left[ \sigma ^{n}(s)\rho (s)%
\right] ,
\end{equation}
where $B_{n}$ is a normalizing constant and the weight function
$\rho(s)$ must be satisfied the case
\begin{equation}
\label{eq5} \frac{d}{ds}[\sigma(s) \rho(s)]=\tau(s) \rho(s).
\end{equation}
On the other hand, the energy eigenvalues of Eq.(\ref{eq4}) are
determined by
\begin{equation}
\label{eq8} \lambda =\lambda _{n}=-n\tau ^{\prime
}-\frac{n(n-1)}{2}\sigma ^{\prime \prime },~~~(n=0,1,2,...).
\end{equation}
Hence, the solution of second-order differential equations can be
solved by means of this method and found their energy eigenvalues
and associated wavefunctions analytically.

\section{Calculations for the Generalized Woods-Saxon Potential}

The interactions between nuclei are commonly described by using a
potential that consist of the Coulomb and the nuclear potentials.
The nuclear potential is usually taken in the Woods-Saxon
potential form, which is one of the important potentials in
nuclear physics. Moreover, the Woods-Saxon potential can be also
used to describe the interaction of a nucleon with a heavy
nucleus. In this paper, we selected a most general form of the
Woods-Saxon potential given by \cite {ref19};
\begin{equation}
\label{eq10} V(x)=-\frac{V_{0}}{1+z}
-\frac{(V_0/a)z}{\left(1+z\right)^2}~,
\end{equation}
where $z=exp\left(\frac{x-R_{0}}{a}\right)$, $V_0$ is the
potential depth, $R_{0}$ is the width of the potential or the
nuclear radius and the parameter $a$ is the thickness of a surface
layer in which the potential falls off from $V=0$ outside to
$V=-V_0$ inside the nucleus. The second term of the right-side of
Eq.(\ref{eq10}) denotes the derivative of Woods-Saxon potential
responsible for the generalization while the first term of the
right-side of the same equation represents the Woods-Saxon
potential. In order to calculate the energy eigenvalues and the
corresponding eigenfunctions, the potential function given by
Eq.(\ref{eq10}) is substituted into the one-dimensional form of
the Schr\"odinger equation:
\begin{equation}
\label{eq11}
\frac{d^2\psi(x)}{dx^2}+\frac{2m}{\hbar^2}\left[E+\frac{V_{0}}{1+qe^{2\alpha
x}}+\frac{Ce^{2\alpha x}}{\left(1+qe^{2\alpha
x}\right)^2}\right]\psi(x)=0.
\end{equation}
Here, some assignments are made in the Schr\"odinger equation such
as $1/a\equiv 2\alpha$, $q=exp(-2\alpha R_0)$ and $C=2\alpha
V_0q$.

We used the NU method to obtain the exact solutions of the SE with
the PT-/non-PT-symmetric potentials only for the s-states. The
energy eigenvalues and eigenfunctions are found in the real or
complex forms and in terms of Jacobi polynomials, respectively.

\subsection{PT-symmetric and non-Hermitian Woods-Saxon case}

We are going to consider different forms of the generalized
Woods-Saxon potential, namely at least one of the parameters is
imaginary. For a special case, we take the potential parameters in
Eq.(\ref{eq10}) as $V_0\rightarrow V_0$ and $\alpha\rightarrow
i\alpha_I$, where $\alpha_I$ is a real parameter of the imaginary
part. Such a potential is called as $PT$-symmetric and also
non-Hermitian, since the property $V(-x)^*=V(x)$ is exist. Hence,
the concept of $PT$-symmetry can be also used in one-dimensional
quantum mechanical problems as many problems exhibiting
$PT$-symmetry \cite {ref23, ref24}. In this case, the new shape of
the potential in one-dimensional space becomes
\begin{equation}
\label{eq} V(x)=-\frac{V_{0}}{1+qe^{2i\alpha_I
x}}-\frac{Ce^{2i\alpha_I x}}{\left(1+qe^{2i\alpha_I
x}\right)^2}
\end{equation}
or it can be written as a complex function
\begin{eqnarray}
\label{eq12} V(x)=-V_0 \left( {\frac{1+q\cos 2\alpha _I x-iq\sin
2\alpha _I x}{1+q^2+2q\cos 2\alpha _I x}} \right)\nonumber\\
~~~~~~~~~~~~~~-C\left({\frac{2q+(1+q^2)\cos 2\alpha _I
x-i(q^2-1)\sin 2\alpha _I x}{\left( {1+q^2+2q\cos 2\alpha _I x}
\right)^2}} \right).
\end{eqnarray}
The type of this potential is known as a complex periodic
potential having $PT$-symmetric and its form is given by $V(x)=i
$sin$^{2n+1}(x)$, $(n=0, 1, 2,...)$ \cite {bender}. A detailed
discussion exhibits for this potential that it has real band
spectra from Ref.[25]. In our case, we will consider the form
given in Eq.(\ref{eq}), following a procedure similar to the
previous section.

Now, in order to apply the NU-method, we rewrite Eq.(\ref{eq11})
by using a new variable of the form $s=-e^{2i\alpha_I x}$,
\begin{equation}
\label{eq13}
\frac{d^2\psi(s)}{ds^2}+\frac{1}{s}\frac{d\psi(s)}{ds}-\frac{m}{2\hbar^2\alpha_I^2s^2}\left[E+\frac{V_{0}}{(1-qs)}-\frac{Cs}{\left(1-qs\right)^2}\right]\psi(s)=0.
\end{equation}
By introducing the following dimensionless parameters
\begin{equation}
\label{eq14}
\varepsilon=-\frac{mE}{2\hbar^2\alpha_I^2}>0~~~(E<0),~~~\beta=\frac{mV_0}{2\hbar^2\alpha_I^2}~~~(\beta>0),~~~~\gamma=\frac{mC}{2\hbar^2\alpha_I^2}~~~(\gamma>0)
\end{equation}
which leads to a hypergeometric type equation defined in
Eq.(\ref{eq1}):
\begin{eqnarray}
\label{eq15}
\frac{d^2\psi(s)}{ds^2}+\frac{1-qs}{s(1-qs)}\frac{d\psi(s)}{ds}+\frac{1}{s^2(1-qs)^2}\left[\varepsilon
(1-qs)^2-\beta (1-qs)+\gamma s\right]\psi(s)=0.
\end{eqnarray}
After the comparison of Eq.(\ref{eq15}) with Eq.(\ref{eq1}), we
obtain the corresponding polynomials as
\begin{equation}
\label{eq16} \stackrel{\sim}{\tau
}(s)=1-qs,~~~{\sigma}(s)=s(1-qs),~~~\stackrel{\sim}{\sigma
}(s)=\varepsilon q^2 s^2-(2\varepsilon q-\beta
q-\gamma)s-\beta+\varepsilon.
\end{equation}
Substituting these polynomials into Eq.(\ref{eq6}), we obtain
$\pi(s)$ function as
\begin{equation}
\label{eq17} \pi (s)=-\frac{qs}{2}\pm
\frac{1}{2}\sqrt{\left(q^2-4\varepsilon
q^2-4kq\right)s^2+4\left(-\beta q-\gamma+2\varepsilon
q+k\right)s+4\left(\beta-\varepsilon\right)}
\end{equation}
taking $ \sigma ^{\prime }(s)=1-2qs$. The discriminant of the
upper expression under the square root has to be zero. Hence, the
expression becomes the square of a polynomial of first degree;
\begin{equation}
\label{eq18} \left(-\beta q-\gamma+2\varepsilon
q+k\right)^2-(\beta-\varepsilon)\left(q^2-4\varepsilon
q^2-4kq\right)=0.
\end{equation}
When the required arrangements are done with respect to the
constant $k$, its double roots are derived as
$k_{1,2}=(\gamma-\beta q)\pm
q\sqrt{\left(\beta-\varepsilon\right)(1-\frac{4\gamma}{q})}$.

Thus substituting, these values for each \textit{k}~into
Eq.(\ref{eq17}) following possible solution is obtained for $\pi
(s)$
\begin{equation}
\label{eq19} \pi(s) = -\frac{qs}{2}\pm
\frac{1}{2}\left\{\begin{array}{ccc}
\left[(2\sqrt{\beta-\varepsilon}+\sqrt{1-\frac{4\gamma}{q}}~)qs-2\sqrt{\beta-\varepsilon}~\right],\nonumber\\
\hskip 0.2cm \mbox{for} \hskip 0.5 cm k=(\gamma-\beta
q)-q\sqrt{(\beta-\varepsilon)(1-\frac{4\gamma}{q})}\\ \\
\left[(2\sqrt{\beta-\varepsilon}-\sqrt{1-\frac{4\gamma}{q}}~)qs-2\sqrt{\beta-\varepsilon}~\right],\nonumber\\
\hskip 0.2cm \mbox{for} \hskip 0.5 cm k=(\gamma-\beta
q)+q\sqrt{(\beta-\varepsilon)(1-\frac{4\gamma}{q})}\\
\end{array}\right.
\end{equation}
After appropriate choice of the polynomial $\pi (s)$ and $k$, we
can write the function $\tau (s)$ which has a negative derivative
as follows
\begin{equation}
\label{eq20} \tau(s)=1+2\sqrt{\beta-\varepsilon}-qs\left(2+
2\sqrt{\beta-\varepsilon}+\sqrt{1-4\gamma/q}~)~\right),\nonumber
\end{equation}
and then its negative derivatives become
\begin{equation}
\label{eq21}
\tau^{\prime}(s)=-q\left(2+2\sqrt{\beta-\varepsilon}+\sqrt{1-4\gamma/q}~\right).
\end{equation}
A particularly interesting result of Eq.(\ref{eq21}) is that the
polynomial $\tau(s)$ is a generalization of the NU method to the
complex quantum mechanics. Therefore, from Eq.(\ref{eq8}) and
Eq.(\ref{eq21}), we write
\begin{equation}
\label{eq22}
\lambda=\lambda_n=nq\left(2+2\sqrt{\beta-\varepsilon}+\sqrt{1-4\gamma/q}~\right)+n(n-1)q,
\end{equation}
and also obtain
\begin{equation}
\label{eq23} \lambda=(\gamma-\beta
q)-q\sqrt{(\beta-\varepsilon)(1-4\gamma/q)}-\frac{q}{2}-
q\left(\sqrt{\beta-\varepsilon}+\frac{1}{2}\sqrt{1-4\gamma/q}\right).
\end{equation}
With the comparison of $\lambda$ values in Eq.(\ref{eq22}) and
Eq.(\ref{eq23}), we found the energy eigenvalues as follows
\begin{eqnarray}
\label{eq24} E_{nq} =\frac{2\hbar
^2}{m}\left(\left(\frac{\alpha_I}{4}\right)^2\left(1+2n+\sqrt{1-2mC/q\hbar
^2}\right)^2+\left(\frac{mV_0/2\hbar^2
\alpha_I}{1+2n+\sqrt{1-2mC/q\hbar
^2}}\right)^2-\frac{mV_0}{4\hbar^2} \right)
\end{eqnarray}
and
\begin{equation}
\label{eq25}
\begin{array}{l}
\varepsilon _{nq} =\frac{\beta
}{2}-\frac{1}{16}\left(1+2n+\sqrt{1-4\gamma
/q}\right)^2-\left(\frac{\beta}{1+2n+\sqrt{1-4\gamma /q}}\right)^2.
\end{array}
\end{equation}
It is clear that the energy eigenvalues have a real part and then
we can say that the real part of energy eigenvalues in
Eq.(\ref{eq25}) determines the energy spectrum in frame of
$PT$-symmetric quantum mechanics. This situation can be also seen
from the condition $\varepsilon_{nq}>0$, since the energy
eigenvalues of the generalized Woods-Saxon potential are negative.
In this sense, the number of discrete levels is finite and
determined by the inequality
$\left(\frac{b}{4}\right)^2+\left(\frac{\beta}{b}\right)^2<\frac{\beta}{2}$,
where $b=1+2n+\sqrt{1-4\gamma /q}$,
$\beta=\frac{mV_{0}}{2\hbar^2\alpha_I^2}$ and
$\gamma=\frac{mC}{2\hbar^2\alpha_I^2}$. Hence, we can write a
condition on the discrete levels for the negative energy spectrum
of the complex parameter generalized Woods-Saxon potential if
$n<\sqrt{mV_{0}/2\hbar^2\alpha_I^2}-\sqrt{1/4-\gamma/q}-1/2$.

Let us now find the corresponding eigenfunctions. The polynomial
solutions of the hypergeometric function $y(s)$ depend on the
determination of weight function $\rho(s)$ which is satisfies the
differential equation $[\sigma (s)\rho (s) ]^{\prime }=\tau
(s)\rho (s)$. Thus, $\rho(s)$ is calculated as
\begin{equation}
\label{eq26}
\rho(s)=\left(1-qs\right)^{\nu-1}s^{2\sqrt{\beta-\varepsilon}},
\end{equation}
where $\nu=1+ \sqrt{1-\frac{4\gamma}{q}}$. Substituting into the
Rodrigues relation given in Eq.(\ref{eq4}), the eigenfunctions are
obtained in the following form
\begin{equation}
\label{eq27}
y_{nq}(s)=A_{n}\left(1-qs\right)^{-(\nu-1)}s^{-2\sqrt{\beta-\varepsilon}}\frac{d^{n}}{ds^{n}}\left[\left(1-qs\right)^{n+\nu-1}s^{n+2\sqrt{\beta-\varepsilon}}%
\right],
\end{equation}
where $A_n=1/n!$. Choosing $q=1$, the polynomial solutions of
$y_n(s)$ are expressed in terms of Jacobi Polynomials
[constant]$P_n^{( 2\sqrt{\beta-\varepsilon},~\nu-1)}(1-2s)$ \cite
{ref21}, which is one of the orthogonal polynomials. By
substituting $\pi(s)$ and $\sigma(s)$ into the expression $\phi
(s)^{\prime }/\phi (s)=\pi (s)/\sigma (s)$, the other part of the
wave function is found as
\begin{equation}
\label{eq28} \phi(s)=(1-s)^{\mu}s^{\sqrt{\beta-\varepsilon}},
\end{equation}
where $\mu=\nu/2$. Combining the Jacobi polynomials with
$\phi(s)$, the s$-$wave complex functions are constructed as
\begin{equation}
\label{eq29}
\psi_n(s)=B_ns^{\sqrt{\beta-\varepsilon}}(1-s)^{\mu}P_n^{(2\sqrt{\beta-\varepsilon},~\nu-1)}(1-2s),
\end{equation}
where $B_{n}$ is the normalizing constant.

\subsection{Non-PT symmetric and non-Hermitian Woods-Saxon case}

In order to be more specific, we are going to take the potential
parameters given in Eq.(\ref{eq10}) as $V_0\rightarrow
V_{0R}+iV_{0I}$ and $\alpha\rightarrow \alpha$, where $V_{0R}$ and
$V_{0I}$ are the real parameters of the complex part. Such a
potential is called as non-$PT$-symmetric but non-Hermitian. In
this case, the condition of $PT$-symmetry is not occurred due to
the fact that $V(-x)^*\neq V(x)$. Hence, its shape becomes
\begin{equation}
\label{eq30}
V(x)=-\left[\frac{V_{0R}}{1+z}+\frac{Cz}{\left(1+z\right)^2}+i\frac{V_{0I}}{1+z}\right]~,
\end{equation}
and these types potentials are called as the pseudo-Hermitian. To
simplify the form of the above equation, we take again an
independent variable by choosing the form $s=-e^{2\alpha x}$, and
then we obtain the generalized equation of hypergeometric type,
\begin{eqnarray}
\label{eq31}
\frac{d^2\psi(s)}{ds^2}+\frac{1-qs}{s(1-qs)}\frac{d\psi(s)}{ds}+\frac{1}{s^2(1-qs)^2}\left[-\varepsilon
(1-qs)^2+(\beta +i\delta)(1-qs)-\gamma s\right]\psi(s)=0,
\end{eqnarray}
for which
\begin{eqnarray}
\label{eq32}
\stackrel{\sim}{\tau}(s)=1-qs,~~~~~{\sigma}(s)=s(1-qs),~~~~\stackrel{\sim}{\sigma}(s)=-\varepsilon
q^2 s^2+(2\varepsilon q-\beta q-i\delta q
-\gamma)s+\beta+i\delta-\varepsilon,\nonumber\\
\varepsilon=-\frac{mE}{2\hbar^2\alpha^2}>0,~~~~\beta=\frac{mV_{0R}}{2\hbar^2\alpha^2},~~~~\gamma=\frac{mC}{2\hbar^2\alpha^2},
~~~~\delta=\frac{mV_{0I}}{2\hbar^2\alpha^2}~~~(\beta,~\gamma,~\delta~>0).~~~~~~
\end{eqnarray}
Following the solution procedures of the NU$-$method, we can
derive the possible solutions for $\pi(s)$ as below
\begin{equation}
\label{eq33}
\pi(s)=-\frac{qs}{2}\pm\frac{1}{2}\left[\left(2\sqrt{\varepsilon-(\beta+i\delta)}\mp\sqrt{1+\frac{4\gamma}{q}}~\right)qs\pm2\sqrt{\varepsilon-(\beta+i\delta)}~\right],
\end{equation}
for $k=(\beta q-\gamma+i\delta q )\pm
q\sqrt{(\varepsilon-\beta-i\delta)(1+\frac{4\gamma}{q})}$. After
performing an appropriate choice for $k$ and $\pi(s)$, we can
write $\tau(s)$ and $\tau^{\prime}$ as
\begin{eqnarray}
\label{eq34}
\tau(s)=1+2\sqrt{\varepsilon-(\beta+i\delta)}-qs\left(2+2\sqrt{\varepsilon-(\beta+i\delta)}~+\sqrt{1+4\gamma/q}\right)\nonumber,\\
\tau^{\prime}(s)=-q\left(2+2\sqrt{\varepsilon-(\beta+i\delta)}~+\sqrt{1+4\gamma/q}\right).
\end{eqnarray}
In the present case, the eigenvalue equation given in
Eq.(\ref{eq8}) is established as follows
\begin{equation}
\label{eq351}
\lambda=\lambda_n=nq\left(2(1+\sqrt{\varepsilon-(\beta+i\delta)}~)+\sqrt{1+\frac{4\gamma}{q}}~\right)+n(n-1)q,
\end{equation}
and then if it is compared with the another form of $\lambda$,
\begin{eqnarray}
\label{eq352} \lambda=(\beta q-\gamma+i\delta q )-
q\sqrt{\left(\varepsilon-\beta-i\delta\right)\left(1+\frac{4\gamma}{q}\right)}-\frac{q}{2}\nonumber\\
~~~~~~~~~~~~~~-\frac{q}{2}\left(2\sqrt{\varepsilon-\beta-i\delta}+\sqrt{1+\frac{4\gamma}{q}}~\right),
\end{eqnarray}
the energy eigenvalues are reduced to the following form
\begin{eqnarray}
\label{eq36}
E_{nq}=-\frac{\hbar^2}{2ma^2}\left\{\frac{1}{16}\left[\sqrt{1+\frac{8mCa^2}{\hbar^2q}}+(1+2n)\right]^2+\frac{4\left(\frac{ma^2}{\hbar^2}\right)^2(V_{0R}^2-V_{0I}^2)}
{\left[\sqrt{1+\frac{8mCa^2}{\hbar^2q}}+(1+2n)\right]^2}\right\}\nonumber\\
-i\frac{\hbar^2}{2ma^2}\left\{\frac{4\left(\frac{ma^2}{\hbar^2}\right)^2V_{0R}V_{0I}}{\left[\sqrt{1+\frac{8mCa^2}{\hbar^2q}}+(1+2n)\right]^2}+\frac{mV_{0I}a^2}{\hbar^2}\right\}.~~~~~~~~~~~~~~~~~~
\end{eqnarray}
In order to mention from an acceptable energy spectra in
Eq.(\ref{eq36}), we have to provide the visible condition given by
if and only if $n>8\left(\delta^2-\beta^2\right)^{1/4}-\kappa/2$ for
both $V_{0R}>V_{0I}$ and $V_{0R}<V_{0I}$, where
$\kappa=\sqrt{1+\frac{4\gamma}{q}~}+1$.

Now, our procedure is that the corresponding unnormalized wave
functions should be determined in terms of Jacobi polynomials. By
considering the Eq.(\ref{eq5}) and using Eq.(\ref{eq34}), we obtain
\begin{equation}
\label{eq37}
\rho(s)=\left(1-qs\right)^{\kappa-1}s^{2\sqrt{\varepsilon-\beta-i\delta}}.
\end{equation}
After that, we derive the polynomial $\phi(s)$ from the equality
$~\phi (s)^{\prime }/\phi (s)=\pi (s)/\sigma (s)$ as
follows
\begin{equation}
\label{eq38}
\phi(s)=s^{\sqrt{\varepsilon-\beta-i\delta}}(1-qs)^{\kappa/2}.
\end{equation}
Hence, we found the relevant wave functions in terms of Jacobi
polynomials
\begin{equation} \label{eq39}
\psi_n(s)=D_ns^{\sqrt{\varepsilon-\beta-i\delta}}(1-s)^{\kappa/2}P_n^{(2\sqrt{\varepsilon-\beta-i\delta},~\kappa-1)}(1-2s),
\end{equation}
where $D_n$ is the another normalizing constant and the parameter
$q$ is chosen as 1.

\section{Conclusions}
The exact solutions of the Schr\"odinger equation for the
generalized Woods-Saxon potential with zero angular momentum are
obtained to found the energy eigenvalues and eigenfunctions.
Nikiforov-Uvarov method is used to solve the Schr\"odinger
equation. The corresponding wave functions are expressed in terms
of Jacobi polynomials. It is pointed out that the NU method can be
generalized by the complex polynomials $\pi(s)$ and $\tau(s)$ with
the framework of PT-symmetric quantum mechanics. When the
potential parameter $\alpha$ is purely complex, it is seen that
the energy eigenvalues are in the real form. In this point, the
shape of potential takes the complex $PT$-symmetric form. It is
not difficult to see that the energy eigenvalues determines the
discrete levels or energy spectrum under the condition of
$n<\sqrt{mV_{0}/2\hbar^2\alpha_I^2}-\sqrt{1/4-\gamma/q}-1/2$.
Therefore, if the parameter of potential $V_0$ is chosen as
complex, it is clear that the energy eigenvalues represent the
complex energy spectrum. We realize that the energy eigenvalues
consist of essentially a real component. Consequently, we find
that the energy eigenvalues of a single particle within the
complex PT and non-PT-symmetric potentials are given in
Eq.(\ref{eq24}) and Eq.(\ref{eq36}), respectively.\\

\end{document}